\documentclass[sigchi-a]{acmart}

\def\BibTeX{{\rm B\kern-.05em{\sc i\kern-.025em b}\kern-.08emT\kern-.1667em\lower.7ex\hbox{E}\kern-.125emX}}
    
\acmYear{2019} 
\acmConference[CHI 2019]{CHI Conference on Human Factors in Computing Systems Proceedings}{May 4--9, 2019}{Glasgow, Scotland Uk}

\begin{document}

%
\title{Interaction with Ubiquitous Robots and Autonomous IoT}

%
\author{Lawrence H. Kim}
\email{lawkim@stanford.edu}
\affiliation{%
  \institution{Stanford University}
  \city{Stanford}
  \state{CA}
  \postcode{94305}
}

\author{Sean Follmer}
\email{sfollmer@stanford.edu}
\affiliation{%
  \institution{Stanford University}
  \city{Stanford}
  \state{CA}
  \postcode{94305}
}

%

%
\begin{abstract}
Robotics have been slowly permeating Internet of Things (IoT) where the previously ubiquitous but static sensors are now given the power to actively navigate the environment and even interact with users. Emergence of these ubiquitous swarms of robots not only opens up the range of possible applications, but also increases the number of elements to study and design for. We do not yet understand how, when, and where these robots should move, manipulate, and touch around people. Through user-centered studies, we aim to better understand how to best design for interaction with Autonomous IoT or a swarm of ubiquitous robots. 
\end{abstract}
%

%
\maketitle

\section{Ubiquitous Robots and Autonomous IoT}
Internet of Things (IoT) has impacted how we live and how we interact with our previously not "smart" objects. Instead of having to physically approach and press buttons to activate a device, we now use our smartphones from wherever we are to easily monitor the current states, command actions, and plan future actions. 

On the other hand, robots have slowly begun to make their ways into our everyday lives. Instead of being restricted to a closed area, robots are now in an open public space delivering food and packages to us. As technology progresses and cost of microchips and robots decreases, the number of robots we will see and interact with can only increase while the robot size will decrease even further. With these swarms of micro-robots, robots will soon become 'invisible', ubiquitous, and truly embedded in our environment. 

We see great possibilities between the intersection of IoT and robots that will lead to autonomous IoT (A-IoT) or ubiquitous robots. With their mobility, ubiquitous robots or A-IoT will enable new types of interaction with people. Two main interactions that are unique and distinct from traditional IoT are their movements and physical interactions where they can come to you. Using user-centered approaches such as elicitation and perception studies, we seek to better understand how to design these ubiquitous robots with people in mind.

\begin{marginfigure}
  \centering
  \includegraphics[width=\linewidth]{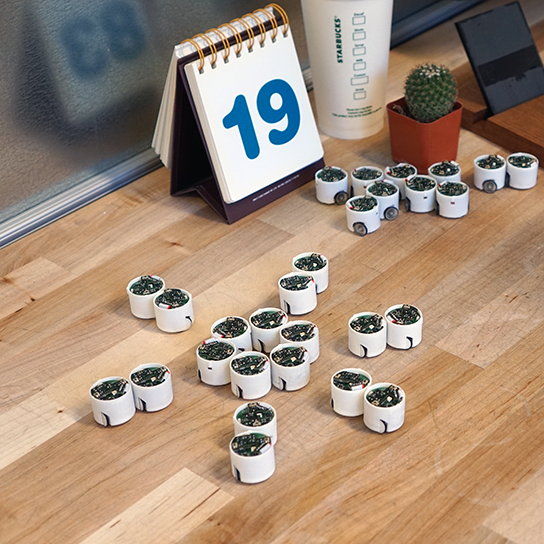}
  \caption{Using iconic patterns to display weather.}
  \label{fig:Iconic}
\end{marginfigure}

\section{Swarm Robot Platform}
\begin{marginfigure}
  \centering
  \includegraphics[width=\linewidth]{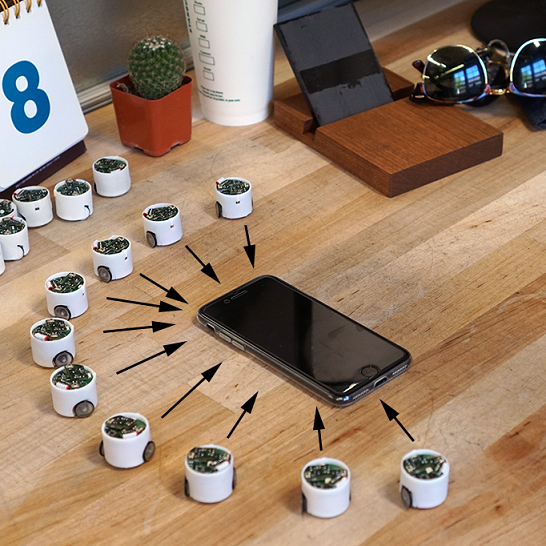}
  \caption{Using abstract swarm motion for phone call notification.}
  \label{fig:UbiSwarm}
\end{marginfigure}

\begin{marginfigure}
  \centering
  \includegraphics[width=\linewidth]{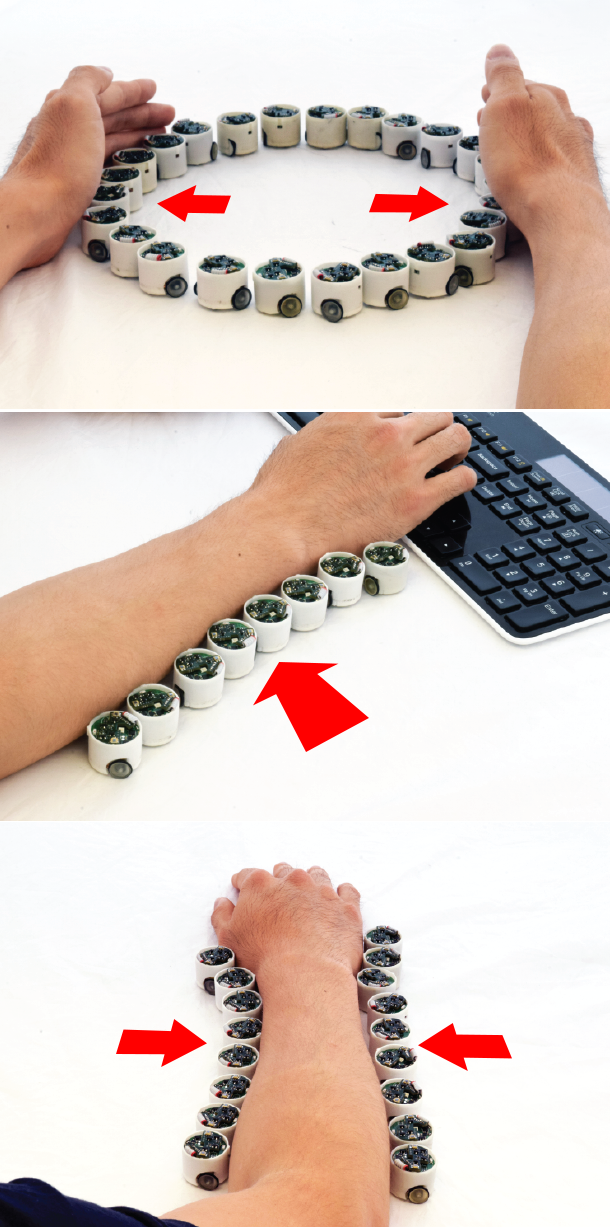}
  \caption{Swarm of robots is used to display various haptics patterns to diferent body parts that are on a surface. It can be used to convey notifcations, social touch, directional cues, etc..}
  \label{fig:SwarmHaptics}
\end{marginfigure}

To enable more realistic user studies with physical robots, we designed and developed Zooids, a platform for Swarm User Interface \cite{le2016zooids}, as shown in Fig. \ref{fig:Iconic}-\ref{fig:SwarmHaptics}. These robots have sensors to navigate the environment, detect user's touch, and measure motion through inertial measurement unit (IMU). While each robot contains its own distributed low-level control loop, a central computer is used to determine and communicate high-level goals to allow fast and real-time control of the robots for user studies. While the Zooids platform enable us to conduct studies with swarm of robots, it has limitations in terms of its nonholonomic drive mechanism and its fixed desktop size. In the future, we hope to build robots with omni-directional drive to enable smoother motion and haptic output, and robots of different sizes and forms to explore their effects on human perception and collaboration.

\section{Swarm Motion as a Display}
One of the central questions for ubiquitous robots or A-IoT is how to display meaningful information to users through robotic motion. Most prior work focus on the use of anthropomorphic features like limbs and facial expressions to convey intent, affect, and information \cite{castellano2007recognising}, but this limits the design space of the robots and their ability to seamlessly blend into our environments. In addition, with the current technology, having anthropomorphic features does not scale well when interacting with many small robots.

There are two ways that the robots can use to display meaningful information. Similar to how we display information in graphical user interface, the robots can form iconic patterns \cite{alonso2011multi} to communicate symbolically as shown in Fig. \ref{fig:Iconic}. We also believe that abstract motion is an important direction to explore. Designers can use essential features like motion and form to convey meaningful information and seamlessly transition between manipulation and display as shown in Fig. \ref{fig:UbiSwarm}. In this paper, we focused on the abstract motion of ubiquitous robots and its effect on human perception in terms of not only emotion, but also user experience, measures for Human-Robot Interaction (HRI), and urgency.

For the study, we varied three multi-robot motion parameters: bio-inspired behavior, speed, and smoothness. From a crowdsourced video user study, we found that the bio-inspired behaviors have significant effects even to a greater extent than that of speed or smoothness. This indicates that when a swarm of robots are moving, how they move especially the coordination between them will significantly impact how humans perceive them. In the future, we hope to build on this and explore more continuous interaction scenarios. For instance, we are interested in exploring how to design swarm motion such that its perceived visibility can be controlled to adapt to different circumstances. To prevent from distracting the user, the robots could slow down such that they are barely noticeable but when there is an urgent task, robots could speed up to notify the user implicitly.

\section{Haptic Interaction}
The mobility of the robots enables them to physically touch and interact with the users. While some HRI researchers have looked at haptic interaction with a single robot \cite{wada2008robot,shiomi2017does}, none have looked at haptic interaction with a swarm of robots. Thus, we first explored the haptic design space for non-anthropomorphic robots. We looked at parameters for single robot including force parameters and contact location, and then expanded to multiple robots exploring different types of coordination possible. 

Using those parameters, we first ran a perception study to better understand user perception of different swarm haptic stimuli. One of the main findings was that there is a trade-off for number of robots where more robots can provide more arousing and urgent haptic stimuli but at the cost of perceived likeability and safety \cite{Kim2019SwarmHaptics}. On a separate elicitation study, we asked users to generate a set of haptic patterns for social touch. The results revealed that users are able to express their thoughts and emotion in various ways with the robots and that they often relied on the visual motion of the robots to convey contextual information for more abstract social touch. This suggest that the visible motion of swarm robots may provide advantage in providing more contexts over traditional invisible haptic devices. In future studies, we would like to explore the use of swarm robots to provide real-time multi-point haptic feedback for interactive applications as prior work has mainly focused on using robots to provide single point feedback or as grasped puck-like devices.

\section{Motivations for the Workshop}
Most of our work so far investigated one-directional discrete interaction where stimuli of short duration were provided to the user. In the future, we hope to design and study interactions that are more continuous, long-term, bi-directional, and user-centered. From this workshop, we are interested in gaining better understanding of state-of-the-art technologies to sense user's internal states like intent and emotion, algorithms to generate appropriate user-centered response, and best design practices and methods for long-term continuous interaction studies.


%
\bibliographystyle{ACM-Reference-Format}
\bibliography{sample-base}

\end{document}